# An approach to constructing super oscillatory functions[*]

M. Mansuripur[†] and P. K. Jakobsen[‡]

[†]College of Optical Sciences, The University of Arizona, Tucson, Arizona 85721
[‡]Department of Mathematics and Statistics, UIT The Arctic University of Norway, Tromsø, Norway

**Abstract**. A recipe is presented for constructing band-limited superoscillating functions that exhibit arbitrarily high frequencies over arbitrarily long intervals.

**1. Introduction**. Superoscillating functions are known to exhibit local frequencies outside the bandwidth of their Fourier transforms.[1] These functions, which were originally discovered in the context of time symmetry in the quantum measurement process,[2-4] have been extensively studied for their intriguing as well as wide-ranging mathematical properties.[5-12] Several applications of these functions have also emerged in recent years, of which a few representative examples are cited here as references.[13-26]

A large number of applications of superoscillatory functions happen to be in the field of optics and photonics. Huang *et al*[17] have demonstrated the feasibility of optical superoscillations by diffraction from a quasiperiodic array of nanoholes. Makris and Psaltis[18] describe superpositions of Bessel functions (also known as diffraction-free solutions of the Helmholtz equation) that not only exhibit superoscillatory features, but also preserve these features upon propagation along the optical axis. Berry[19] has discussed the circumstances under which subwavelength superoscillatory detail can be exactly reproduced upon nonparaxial transmission. Theoretical analysis of a confocal microscope that uses a special class of superoscillatory functions to produce optimal focused spots was recently reported by Rogers *et al*.[20] These spots, which are formed by a judicious superposition of circular prolate spheroidal wave functions, exhibit bright central regions having subwavelength diameters, a relatively large zone of silence surrounding the central bright spot, and an optimized ratio of maximum intensity outside the zone of silence to the peak intensity at the central spot. Eliezer *et al* have demonstrated super defocusing of light by optical sub-oscillations,[21] and also shown that the temporal resolution limit can be broken by superoscillating optical beats.[22] For a recent review of the applications of superoscillations in optics and image science, see Gbur.[23]

As for mathematical construction of superoscillatory functions, several methods have been proposed. Berry and co-workers[1,5,10,24] have proposed and analyzed a number of functions that exhibit oscillations at arbitrary frequencies beyond the maximum frequency present in the corresponding Fourier spectrum; these oscillations can persist over arbitrarily long intervals. Berry has also discussed the strength of such functions outside their range of superoscillations relative to that of the superoscillations themselves, investigated the effects of noise on the visibility and persistence of superoscillations, and used superoscillatory functions as a basis to represent fast-varying functions (including fractals) by functions of arbitrarily narrow spectral width.[11,25] Qiao[27] has shown that the zeros of a square-integrable and bandlimited waveform can be shifted around arbitrarily, thus providing a mechanism for bringing several zeros close together to create superoscillating waveforms. Construction of quantum mechanical wave-

---

functions containing superoscillatory regions with optimal characteristics has been discussed by Kempf and Ferreira.[28] These authors demonstrate a method of obtaining wave-functions with the most pronounced superoscillations, and point out certain unusual phenomena associated with such wave-functions that are of measurement theoretic, thermodynamic, and information theoretic interest. The fragility of superoscillations in conjunction with the instability of reconstructing oversampled signals was the subject of a study by Ferreira *et al*.[29] Direct construction of superoscillatory functions having optimum energy concentration with optimal numerical stability has been discussed by Lee and Ferreira[30-32]. Katzav *et al*[33,34] have defined the yield of a superoscillatory function as the ratio of its energy within the superoscillating region to the overall energy of the function. They have then constructed yield-optimized superoscillations, and systematically investigated the yield statistics of their functions in the presence of noise in the corresponding Fourier coefficients.

The present paper introduces a class of functions that are fairly easy to construct mathematically and/or numerically, and that exhibit many of the desirable characteristics of superoscillating functions. In some respects, the class of waveforms introduced here is similar to that described by Chojnacki and Kempf,[7] but there exist important differences between the two classes of functions. Specifically, the superoscillating features of the functions described in the present paper are constructed separately and independently of the envelope function that serves to control and to moderate the rapid growth of the superoscillating part of the function outside its region of superoscillations. The general idea behind our method of constructing superoscillatory functions is described in Sec.2, in the context of a truncated Euler expansion of sinusoidal waveforms and a rather simple envelope function. Examples of alternative envelopes with more degrees of freedom and, consequently, more flexibility in selecting their characteristic features, are given in Sec.3. Section 4 introduces the idea of adding zeros to and/or removing zeros from bandlimited envelope functions, thereby providing additional flexibility in the design of super-oscillators. In Sec.5, we present an argument in support of the general observation that the yield of a superoscillating function, defined as its ratio of superoscillating energy to overall energy, declines as the frequency and/or duration of its superoscillations grow. This argument enables us to place an upper bound, albeit a weak one, on the yield ratio of our superoscillating functions. The paper closes with a few concluding remarks in Sec.6.

**2. A recipe for constructing superoscillating functions**. The Euler expansion of $\cos(2\pi f_0 t)$ as an infinite product of first-order polynomials $(1 - t/t_n)$ that reproduce the zeros $t_n = \pm(2n-1)/(4f_0)$ of the cosine function is typically written as follows:[35]

$$\cos(2\pi f_0 t) = \prod_{n=1}^{\infty} \left[1 - \left(\frac{4f_0 t}{2n-1}\right)^2\right]. \qquad (1)$$

If the above Euler product is terminated at $n = N$, the remaining terms retain the first $2N$ zeros of $\cos(2\pi f_0 t)$, i.e., those located between $t = \pm N/(2f_0)$, but the truncated product will grow rapidly (as $|t|^{2N}$) away from the $\pm N^{\text{th}}$ zeros. As a matter of fact, the truncated product provides a reasonable approximation to $\cos(2\pi f_0 t)$ only over the interval $|t| \lesssim \sqrt{N}/(4f_0)$, beyond which the product oscillates wildly even though the zeros of $\cos(2\pi f_0 t)$ are correctly reproduced all the way to $t = \pm N/(2f_0)$. As shown in Appendix A, the truncated Euler product at $N \gg f_0|t|$ is well approximated as

$$\prod_{n=1}^{N}\left[1-\left(\frac{4f_0 t}{2n-1}\right)^2\right] \cong \exp[(2f_0 t)^2/(N+\tfrac{1}{2})]\cos(2\pi f_0 t). \qquad (2)$$

Thus, over the interval $|t| \lesssim \sqrt{N}/(4f_0)$, the amplitude of $\cos(2\pi f_0 t)$ stays between 1.0 and $e^{1/4} \cong 1.28$.

Next, let us consider a band-limited function, such as $g(t) = \mathrm{sinc}(t) = \sin(\pi t)/(\pi t)$, whose Fourier transform $G(f) = \int_{-\infty}^{\infty} g(t)\exp(-i2\pi f t)\,dt = \mathrm{rect}(f)$ is confined to the frequency range $|f| \leq \tfrac{1}{2}$. The function $g^\nu(t)$, where $\nu$ is a large positive integer, is also band-limited, its Fourier spectrum $G_\nu(f)$ being the result of $\nu$ repeated convolutions of $G(f)$ with itself. The Fourier transform $G_\nu(f)$ of the $\nu^{\text{th}}$ power of $g(t)$ is thus seen to be confined to $|f| \leq \nu/2$.

Upon scaling the argument of $g^\nu(t)$, we arrive at the (band-limited) time-domain function $g^\nu(t/\nu)$, whose spectrum $\nu G_\nu(\nu f)$ resides within the frequency range $|f| \leq \tfrac{1}{2}$. For large values of $\nu$, the function $g^\nu(t/\nu)$, which resembles a Gaussian, is fairly smooth and relatively flat over the interval $|t| \lesssim \sqrt{\nu}/\pi$. For instance, in the case of $g(t) = \mathrm{sinc}(t)$, we have $g^\nu(0) = 1$ and

$$g^\nu(t = \sqrt{\nu}/\pi\nu) = \left[\sqrt{\nu}\sin(1/\sqrt{\nu})\right]^\nu = \left[1 - \tfrac{1}{3!}(1/\sqrt{\nu})^2 + \cdots\right]^\nu \cong 1 - \tfrac{1}{6} \cong 0.83. \qquad (3)$$

If we now multiply the truncated Euler product of Eq.(1) with the wide, smooth, Gaussian-like, and bandlimited function $g^\nu(t/\nu)$, we obtain a superoscillating function, as follows:

$$h(t; \nu, f_0, N) = g^\nu(t/\nu) \prod_{n=1}^{N}\left[1-\left(\frac{4f_0 t}{2n-1}\right)^2\right]. \qquad (4)$$

Since, in the above equation, $g^\nu(t/\nu)$ is multiplied by a polynomial of order $2N$ in the time variable $t$, the Fourier transform $H(f; \nu, f_0, N)$ of the product will be a linear superposition of the (band-limited) function $G_\nu(\nu f)$ and its various derivatives with respect to $f$. In addition to the function $G_\nu(\nu f)$, its $2^{\text{nd}}, 4^{\text{th}}, 6^{\text{th}}, \cdots, (2N)^{\text{th}}$ derivatives will appear in the superposition. Needless to say, all these derivatives are also band-limited, since they exist solely within the frequency range $|f| < \tfrac{1}{2}$. At the extreme points $f = \pm\tfrac{1}{2}$ of the spectral range, the derivatives of $G_\nu(\nu f)$ will be well-behaved so long as $\nu > 2N$ (see Appendix B). Violation of this condition would introduce $\delta$-functions and derivatives of $\delta$-functions at $f = \pm\tfrac{1}{2}$ (and possibly elsewhere within the bandwidth), which renders the super-oscillating function non-square-integrable. Stated differently, if $\nu \leq 2N$, the tails of the envelope function $g^\nu(t/\nu)$ fail to decline faster than $|t|^{2N}$, which is a necessary condition if the super-oscillating function is to approach zero as $|t| \to \infty$.

All in all, $h(t; \nu, f_0, N)$ of Eq.(4) is a superoscillating function whose spectrum is limited to $|f| < \tfrac{1}{2}$, mimics the function $\cos(2\pi f_0 t)$ over the interval $|t| \lesssim \sqrt{N}/(4f_0)$, grows rapidly beyond the superoscillating region (perhaps as fast as $|t|^{2N}$ after the $\pm N^{\text{th}}$ zeros of the truncated Euler product), and, provided that $\nu > 2N$, declines to zero as $|t|^{2N-\nu}$ when $|t| \to \infty$.

As a numerical example, suppose $f_0 = 50$ Hz. If we choose $N = 4 \times 10^4$, the first 100 periods of the cosine function (located between $t = \pm 1$ sec) will be well reproduced by the truncated Euler product of Eq.(4). If the value of $\nu$ is now chosen to be greater than $2N = 8 \times 10^4$, the envelope function $g^\nu(t/\nu)$ will be quite flat and smooth (essentially equal to 1) in the interval $|t| < 1$ sec, but, of course, it will have significant values—less than 1.0 but greater than zero—up to, say, $t \cong 500$ sec. The truncated Euler product will grow substantially during the interval $1 < |t| < 500$ sec and beyond, but the eventual rapid decline of the envelope $g^\nu(t/\nu)$ causes the superoscillating function $h(t; \nu, f_0, N)$ to fade as $|t| \to \infty$.

**3. Alternative envelope functions.** Our choice in Sec. 2 of the simple function sinc($t$) for $g(t)$ has been primarily intended for demonstration purposes. There exists a large class of functions whose members could be substituted for $g(t)$ in the preceding discussion. A few examples of such functions are listed below, but many more can be constructed along similar lines of reasoning. Mathematical details pertaining to Fourier transformation of the listed functions are presented in Appendix C.

i) The function $g(t) = 3[\text{sinc}(2t) - \cos(2\pi t)]/(2\pi t)^2$, whose Fourier transform is given by $G(f) = \frac{3}{4}(1 - f^2)\text{rect}(f/2)$, is band-limited within $|f| < 1$. The function peaks at $g(0) = 1$.

ii) The function $g(t) = \sqrt{\pi}\,\Gamma(\kappa + 1)J_{\kappa+\frac{1}{2}}(2\pi t)/(\pi t)^{\kappa+\frac{1}{2}}$, parameterized with the real-valued $\kappa > -1$, has $G(f) = (1 - f^2)^\kappa \text{rect}(f/2)$ for its Fourier transform, which indicates that the function's frequency spectrum is confined to the interval $|f| < 1$. Here, $J_{\kappa+\frac{1}{2}}(\cdot)$ is the Bessel function of first kind, order $\kappa + \frac{1}{2}$, and $\Gamma(\cdot)$ is Euler's gamma function. The function $g(t)$ peaks at $t = 0$, where $g(0) = \sqrt{\pi}\,\Gamma(\kappa + 1)/\Gamma(\kappa + \frac{3}{2})$.

iii) The function $g(t) = \pi\Gamma(\kappa + 1)/[2^\kappa \Gamma(1 + \frac{1}{2}\kappa + \pi t)\Gamma(1 + \frac{1}{2}\kappa - \pi t)]$, parameterized with the real-valued $\kappa > -2$, is bandlimited within the interval $|f| < \pi/2$. Its Fourier transform is $G(f) = \cos^\kappa(f)\,\text{rect}(f/\pi)$. The function peaks at $g(0) = \sqrt{\pi}\,\Gamma(\frac{1}{2} + \frac{1}{2}\kappa)/\Gamma(1 + \frac{1}{2}\kappa)$.

iv) The function $G(f) = \exp\left(\frac{1}{f^2 - 1}\right)\text{rect}(f/2)$ depicted in Fig.1(a) is well-behaved within its bandwidth $|f| < 1$, having smooth transitions to zero at the edges of the band. Although an analytic expression for its inverse Fourier transform $g(t)$ is not known to the authors, the function is readily transformable numerically. A plot of $g(t)$ over the interval $t = -5$ to $5$ is shown in Fig.1(b). *Mathematica*™ can readily compute $g(t)$ over the entire interval $|t| \leq 500$ with 10 digits of precision. At $t = \pm 500$, the magnitude of $g(t)$ is on the order of $10^{-27}$.

Considering that $G(f)$ has smooth and continuous derivatives of all orders within its bandwidth of $|f| < 1$, there will be no need to raise the corresponding $g(t)$ to the power of $\nu > 2N$ to ensure proper behavior for the derivatives of $G(f)$ at the extreme points $f = \pm 1$ of the bandwidth. Numerical calculations up to $t = 10^4$ indicate that the tails of $g(t)$ decay exponentially as $\sim \exp(3.70153 - 3.45055|t|^{0.47})$ when $|t| \to \infty$; this, of course, is faster than any polynomial decay rate, which renders this particular $g(t)$ an ideal envelope function.

Needless to say, the time-domain width of the present $g(t)$ can be adjusted by scaling $G(f)$ as $\eta G(\eta f)$, where $\eta > 0$ is any desired scale factor. Alternatively, one could adjust the general profile of $g(t)$ by selecting different spectral distributions of similar structure, such as $G(f) = \exp[f^{2m}/(f^{2n} - 1)]\text{rect}(f/2)$, where $m$ and $n$ are arbitrary positive integers.

Figure 2(a) shows a typical plot of the function $h(t; \nu, f_o, N)$ for $\nu = 5$, $f_o = 50$ Hz, and $N = 10$; the spectrum of the envelope function used here is $G(f) = \exp\left(\frac{1}{f^2 - 1}\right)\text{rect}(f/2)$. The function's superoscillatory region is located in its midsection, which is too weak to be seen on the scale of Fig.2(a), where $t_{\max} = 80$ sec and $h_{\max} \cong 4 \times 10^{40}$. A magnified view of the midsection of the function appears in Fig.2(b), where the amplitude at $t = 0$ equals 0.0165. The superoscillating region can be broadened by raising the value of $N$ to, say, 100, as depicted in Fig.2(c). This, of course, will cause the maximum amplitude and overall width of the function to grow enormously, which, although manageable by existing computers, does not produce pretty pictures.

A wide range of the parameters $\nu$, $f_o$, and $N$ can be used to numerically evaluate the function $h(t; \nu, f_o, N)$ of Eq.(4) over different intervals of $t$. Although the various envelope functions described in Sections 2 and 3 yield accurate and stable solutions for small to moderate values of $N$, we have encountered numerical difficulties with larger values of $N$, primarily due to the requirement that $\nu$ be greater than $2N$. The exception, of course, has been the envelope function having the infinitely-smooth spectrum $G(f) = \exp\left(\frac{1}{f^2-1}\right)\text{rect}(f/2)$, which can be used with any value of $\nu$ (including $\nu = 1$) regardless of the value chosen for $N$. The numerical difficulties in this case arise only when the function needs to be evaluated at large values of both $N$ and $|t|$.

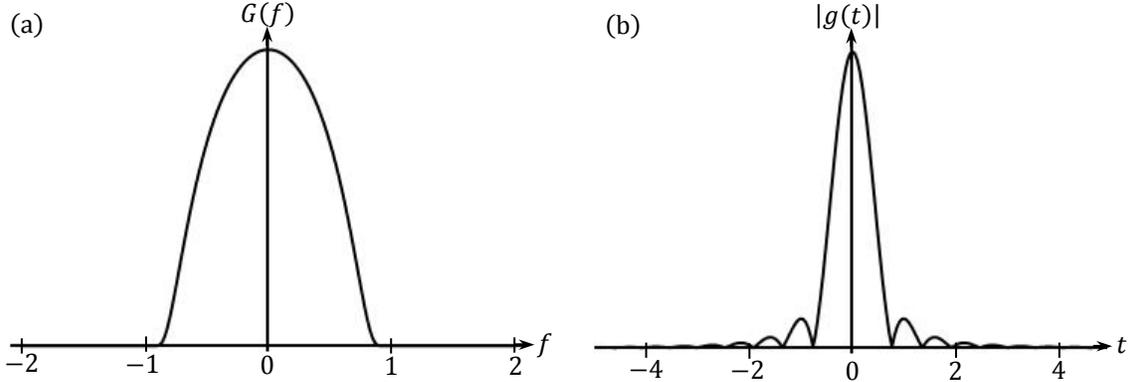

**Fig.1**. Plots of the function $G(f)$ and its inverse Fourier transform $g(t)$. The peak value of $G(f)$ is $e^{-1} \cong 0.368$, while that of $g(t)$ is 0.44.

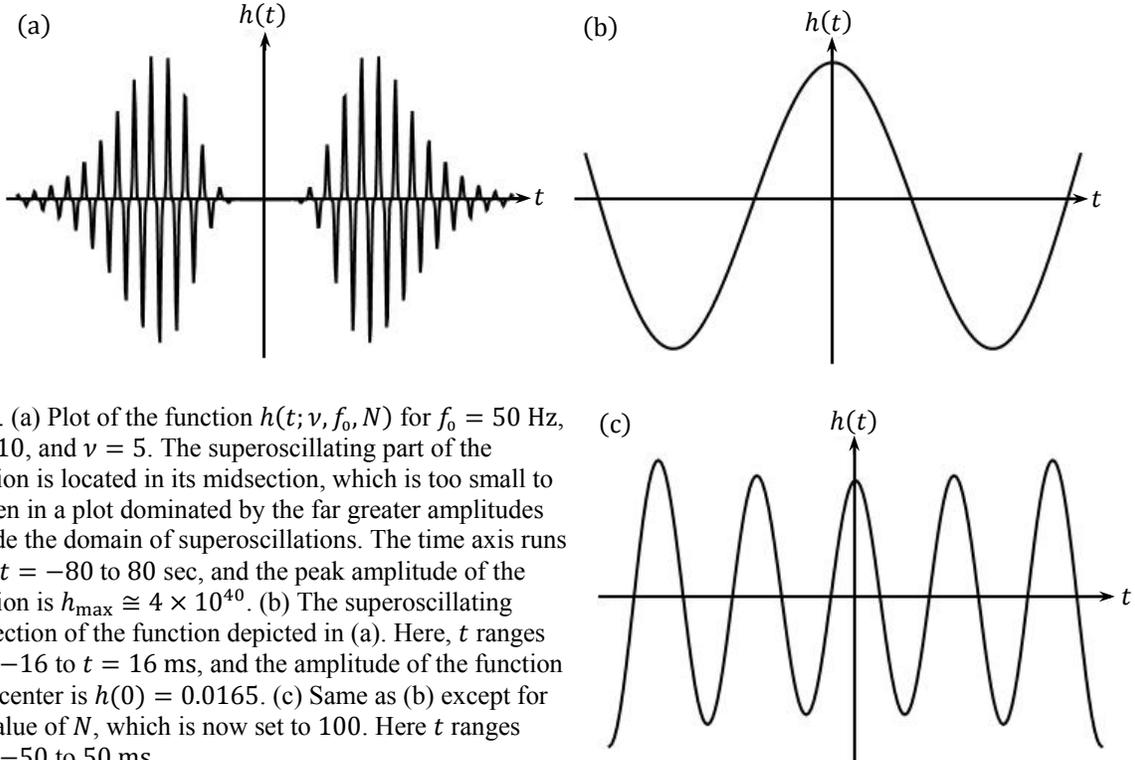

**Fig.2.** (a) Plot of the function $h(t; \nu, f_o, N)$ for $f_o = 50$ Hz, $N = 10$, and $\nu = 5$. The superoscillating part of the function is located in its midsection, which is too small to be seen in a plot dominated by the far greater amplitudes outside the domain of superoscillations. The time axis runs from $t = -80$ to $80$ sec, and the peak amplitude of the function is $h_{\max} \cong 4 \times 10^{40}$. (b) The superoscillating midsection of the function depicted in (a). Here, $t$ ranges from $-16$ to $t = 16$ ms, and the amplitude of the function at its center is $h(0) = 0.0165$. (c) Same as (b) except for the value of $N$, which is now set to 100. Here $t$ ranges from $-50$ to $50$ ms.

**4. Adding zeros to and/or removing zeros from an envelope.** All the envelope functions described in the preceding section can be further modified by removing some of their zeros, and/or by adding new zeros to their profiles. To add a new zero, say, at $t = t_o$, to an existing envelope $g(t)$, simply multiply $g(t)$ with $1 - (t/t_o)$. As discussed earlier, the spectrum $G(f) - (i/2\pi t_o)G'(f)$ of the product function will have a bandwidth no greater than that of $G(f)$. Moreover, the product function remains square-integrable if $G(f)$ is continuous and square-integrable, in which case $G'(f)$ will not contain any delta-functions. Stated differently, the tails of $g(t)$ must decline at least as fast as $1/|t|^2$ when $|t| \to \infty$ if the product function $[1 - (t/t_o)]g(t)$ is to remain square-integrable. (Note: in the special case of $t_o = 0$, the product function and its Fourier transform will be $tg(t)$ and $(i/2\pi)G'(f)$, respectively.) Needless to say, any number of zeros at any number of desired locations can be introduced into a bandlimited envelope $g(t)$, provided that $g(t)$ approaches zero sufficiently rapidly when $|t| \to \infty$.

Removing one or more zeros from a bandlimited and square-integrable envelope $g(t)$ is also permissible. To remove a zero of $g(t)$, say, one located at $t = t_1$, simply divide $g(t)$ by $1 - (t/t_1)$. For reasons that are explained in Appendix D, this operation does *not* result in an increase of the bandwidth, meaning that the bandwidth of $g(t)/[1 - (t/t_1)]$ will be no greater than that of $g(t)$. If $t_1$ happens to be an $n^{\text{th}}$ order zero of $g(t)$, then it must be removed by repeated application of the same procedure, namely, by dividing $g(t)$ by $[1 - (t/t_1)]^n$. It is clear that removing one or more of the zeros of an envelope function will cause its tails to decline more rapidly with an increasing $|t|$, i.e., in proportion to $|t|$ raised to the power of the number of zeros that have been removed.

A special case of adding and removing zeros involves the substitution of one zero for another, i.e., multiplying $g(t)$ with $(t - t_o)/(t - t_1)$. This case has been discussed by Qiao, and referred to as the zero-shifting principle for functions in the Paley-Wiener space.[27]

**5. A weak upper-bound on the yield of superoscillatory functions.** It is well known that the overall energy $\mathcal{E}_h = \int_{-\infty}^{\infty} h^2(t) dt = \int_{-f_{\max}}^{f_{\max}} |H(f)|^2 df$ of a superoscillating function $h(t)$ increases greatly when its superoscillations grow in amplitude, frequency, or duration. Assuming that $h(t) = a_s \cos(2\pi f_s t)$ during the time interval $(t_o, t_o + \tau_s)$, where $f_s > f_{\max}$, the energy content of the superoscillatory part of $h(t)$ will be $\mathcal{E}_s = \int_{t_o}^{t_o+\tau_s} a_s^2 \cos^2(2\pi f_s t) \, dt = \frac{1}{2} a_s^2 \tau_s$. We assume the host function contains a reasonably large number of superoscillation cycles, that is, $f_s \tau_s \gg 1$. The goal in the present section is to estimate a reasonable upper bound on the ratio $\mathcal{E}_s/\mathcal{E}_h$ of the energy content of these superoscillations to the overall energy of the host function.

Our approach to obtaining an upper bound on the yield of the superoscillatory function $h(t)$ is based on the observation that the passage of $h(t)$ through a linear, shift-invariant filter can be analyzed in two equivalent ways. In the first method, the impulse-response $q(t)$ of the filter is convolved with the input function to yield the output $h(t) * q(t)$ of the filter. In the second method, the transfer function $Q(f)$ of the filter, which is just the Fourier transform of $q(t)$, is multiplied by $H(f)$ to arrive at the Fourier transform $H(f)Q(f)$ of the filter's output. Figure 3 shows a typical plot of $|H(f)|$ for a bandlimited signal, as well as a typical plot of $|Q(f)|$ for a narrowband filter tuned to the superoscillation frequency $f_s$ of the input signal.

When $f_s \gg f_{\max}$ and the filter's bandwidth $\Delta f$ is sufficiently narrow, the tails of $Q(f)$ that overlap with $H(f)$ will be weak, and the total output power $\int_{-f_{\max}}^{f_{\max}} |H(f)Q(f)|^2 df$ will be small. At the same time, during the interval $\tau_s$, a fraction of the superoscillating signal $a_s \cos(2\pi f_s t)$ will pass through the filter due to its convolution with $q(t)$, whose oscillation frequency coincides with the frequency $f_s$ of the superoscillations. The fact that the strength of the output signal obtained by the latter argument cannot exceed that obtained by the former, enables us to estimate an upper bound on the yield ratio $\mathcal{E}_s/\mathcal{E}_h$.

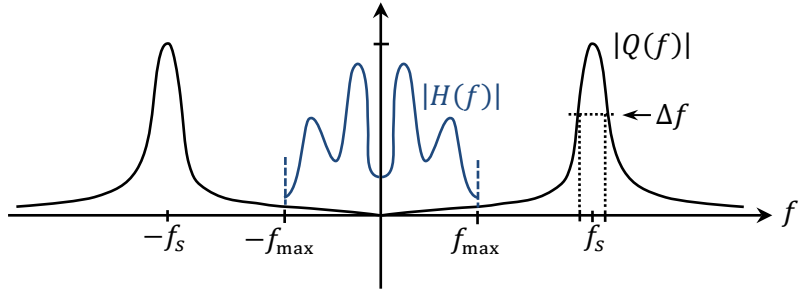

**Fig.3**. A Plot of the magnitude $|Q(f)|$ of the transfer function of the filter, superimposed on the spectral distribution $|H(f)|$ of a superoscillating signal whose frequency content is confined to the interval $|f| \leq f_{\max}$. The filter, which has a relatively narrow bandwidth $\Delta f$, is tuned to the superoscillations frequency $f_s$ of the input signal $h(t)$.

As a simple example, let the impulse-response of a linear, shift-invariant filter having a finite duration $\tau < \tau_s$ be given by

$$q_\tau(t) = \exp\{1/[(2t/\tau)^2 - 1]\}\operatorname{rect}(t/\tau)\cos(2\pi f_s t). \tag{5}$$

Denoting by $\hat{Q}(f)$ the Fourier transform of the function $\exp[1/(t^2 - 1)]\operatorname{rect}(t/2)$, we express the Fourier transform $Q_\tau(f)$ of $q_\tau(t)$ as follows:

$$Q_\tau(f) = (\tau/4)\{\hat{Q}[\tau(f - f_s)/2] + \hat{Q}[\tau(f + f_s)/2]\}. \tag{6}$$

Note that $\hat{Q}(f)$ is the same function as that depicted in Fig.1(b) — except for its variable here being $f$ rather than $t$. The peak value of $\hat{Q}(f)$ is thus given by $\hat{Q}(0) \cong 0.44$. In what follows, we shall ignore the slight overlap between the left and right halves of $Q_\tau(f)$ given by Eq.(6), and set $Q_\tau(\pm f_s) \cong 0.11\tau$. Thus, when a uniform and infinitely-long sinusoidal signal $a_s\cos(2\pi f_s t)$ enters the filter, it emerges at the output as $0.11\tau a_s\cos(2\pi f_s t)$. Considering that the superoscillations have a finite duration $\tau_s$, and that the width of the filter's impulse-response $q_\tau(t)$ is $\tau$, the above sinusoidal output will exist only during a time interval $\tau_s - \tau$ (i.e., when the impulse-response function of width $\tau$ fully overlaps the superoscillations of duration $\tau_s$). The output energy of the filter during this window in time is thus given by

$$\tfrac{1}{2}(0.11\tau a_s)^2(\tau_s - \tau) = 0.0121\tau^2[1 - (\tau/\tau_s)]\mathcal{E}_s. \tag{7}$$

The above energy, of course, cannot exceed the overall energy output of the filter, namely, $\int_{-f_{\max}}^{f_{\max}}|Q_\tau(f)H(f)|^2 df$. Ignoring, once again, the slight overlap between the left and right halves of $Q_\tau(f)$ of Eq.(6), and noting that, within the bandwidth $[-f_{\max}, f_{\max}]$ of the input signal,

$$|Q_\tau(f)| < (\tau/4)|\hat{Q}[\tau(f_s - f_{\max})/2]|, \tag{8}$$

we arrive at

$$0.0121\tau^2[1 - (\tau/\tau_s)]\mathcal{E}_s < (\tau/4)^2\hat{Q}^2[\tau(f_s - f_{\max})/2]\mathcal{E}_h. \tag{9}$$

We now use the aforementioned asymptotic expression $\exp(3.70153 - 3.45055|f|^{0.47})$ for the tails of $\hat{Q}(f)$ to arrive at an upper bound on the yield ratio $\mathcal{E}_s/\mathcal{E}_h$ of $h(t)$, as follows:

$$\frac{\mathcal{E}_s}{\mathcal{E}_h} < \frac{\exp\{9 - 5[\tau(f_s - f_{\max})]^{0.47}\}}{1 - (\tau/\tau_s)}; \qquad (f_s \gg f_{\max} \text{ and } \tau_s f_s \gg 1). \tag{10}$$

The expression on the right-hand side of Eq.(10), a function of the width $\tau$ of the impulse-response $q_\tau(t)$, should be minimized for a tight upper bound on the yield of the superoscillatory function $h(t)$. For instance, given $f_{\max} = 1$, $f_s = 50$, and $\tau_s = 1$, the optimum value of $\tau$ in Eq.(10) is found to be $\tau = 0.934$, resulting in $\mathcal{E}_s/\mathcal{E}_h < 0.979 \times 10^{-8}$.

Due to our conservative choice of the impulse-response function in Eq.(5), the upper bound on the yield ratio $\mathcal{E}_s/\mathcal{E}_h$ obtained in Eq.(10) is relatively weak. The upper bound can be substantially strengthened by a more aggressive choice of the filter, such as one having the Gaussian impulse-response $q_\tau(t) = \exp[-(t/\tau)^2]\cos(2\pi f_s t)$. However, the infinite width of $q_\tau(t)$ in this case demands a more nuanced approach to optimizing the parameter $\tau$ in conjunction with an estimate for the rate of growth of $h(t)$ in the immediate neighborhood of its superoscillations. These issues will take us far beyond the scope of the present paper and are best left for a different discussion in a separate paper.

**6. Concluding remarks**. We close by noting that the Euler product in Eq.(1) does *not* necessarily have to represent a cosine function. In other words, any arbitrary set of zeros can be chosen for the superoscillating part of the function. Also, the envelope function $g^\nu(t/\nu)$ described in Sec. 2 does *not* have to be a single function $g(t)$ raised to some integer power $\nu$, then accordingly scaled. Any compendium of band-limited functions that satisfy the aforementioned minimal criteria (i.e., differentiability with respect to $f$ of the overall Fourier spectrum up to and including the $(2N)^{\text{th}}$ order, flatness of the envelope function over the interval $|t| \lesssim \sqrt{N}/(4f_0)$, and a rate of decline faster than $|t|^{2N}$ as $|t| \to \infty$) can be multiplied together, then properly scaled, to form an acceptable envelope for the superoscillating function.

There exists a large degree of freedom in choosing the superoscillatory part of the function $h(t)$ as well as the corresponding envelope $g(t)$. One can choose the set of parameters $f_0$, $N$, and $\nu$ in conjunction with a desirable envelope function, then examine the behavior of $h(t)$ numerically to see if it is suitable for a specific application. Our numerical calculations confirm that a large number of such functions can be constructed that have the desirable characteristics of high frequency, uniform amplitude, and long duration of superoscillations, together with square-integrability and limited bandwidth for the host function. We do not have any particular insight into the computability of these functions for large values of the relevant parameters, their optimal yield ratios, or their practical applications beyond what is already available in the vast literature of the subject.

**Acknowledgement**: The authors are grateful to Rolf Binder, Isak Kilen, Miro Kolesik, Tobias Mansuripur, and Ewan Wright for their valuable comments and suggestions. This work has been supported in part by the AFOSR grant FA9550-19-1-0032.

## Appendix A

The truncated Euler product expansion of $\cos(2\pi f_0 t)$ can be expressed in terms of the gamma function, $\Gamma(t)$, as follows:

$$\prod_{n=1}^{N}\left[1-\left(\frac{4f_0 t}{2n-1}\right)^2\right] = \prod_{n=1}^{N}\frac{(n-\tfrac{1}{2}-2f_0 t)(n-\tfrac{1}{2}+2f_0 t)}{(n-\tfrac{1}{2})^2}$$

$$= \frac{\Gamma(N+\tfrac{1}{2}-2f_0 t)}{\Gamma(\tfrac{1}{2}-2f_0 t)} \times \frac{\Gamma(N+\tfrac{1}{2}+2f_0 t)}{\Gamma(\tfrac{1}{2}+2f_0 t)} \bigg/ \left[\frac{\Gamma(N+\tfrac{1}{2})}{\Gamma(\tfrac{1}{2})}\right]^2$$

$$= \frac{\Gamma(N+\tfrac{1}{2}-2f_0 t)\times\Gamma(N+\tfrac{1}{2}+2f_0 t)}{[\Gamma(N+\tfrac{1}{2})]^2}\times\cos(2\pi f_0 t). \qquad (A1)$$

In arriving at Eq.(A1), we have invoked the following properties of the gamma function:

$$\Gamma(x+1) = x\Gamma(x); \qquad\qquad (\text{G\&R }\mathbf{8.331}\text{-1}), \qquad (A2a)$$

$$\Gamma(\tfrac{1}{2}-x)\Gamma(\tfrac{1}{2}+x) = \pi/\cos(\pi x); \qquad\qquad (\text{G\&R }\mathbf{8.334}\text{-2}), \qquad (A2b)$$

$$\Gamma(\tfrac{1}{2}) = \sqrt{\pi}; \qquad\qquad (\text{G\&R }\mathbf{8.338}\text{-2}).^{36} \qquad (A2c)$$

For sufficiently large values of $N$ (i.e., $N \gg f_0|t|$), the coefficient of $\cos(2\pi f_0 t)$ in Eq.(A1) can be approximated using $\Gamma(x) \cong \sqrt{2\pi}e^{-x}x^{x-\tfrac{1}{2}}$ for $x \gg 1$ (G&R **8.327**-1)[36], as follows:

$$\frac{\Gamma(N+\tfrac{1}{2}-2f_0 t)\times\Gamma(N+\tfrac{1}{2}+2f_0 t)}{[\Gamma(N+\tfrac{1}{2})]^2} \cong \left(1-\frac{4f_0 t}{2N+1}\right)^{N-2f_0 t}\left(1+\frac{4f_0 t}{2N+1}\right)^{N+2f_0 t}$$

$$= \left[1-\left(\frac{4f_0 t}{2N+1}\right)^2\right]^N \left(1-\frac{4f_0 t}{2N+1}\right)^{-2f_0 t}\left(1+\frac{4f_0 t}{2N+1}\right)^{2f_0 t}$$

$\boxed{1\pm\varepsilon \cong \exp(\pm\varepsilon)} \to \cong \exp\left[-N\left(\frac{4f_0 t}{2N+1}\right)^2\right]\exp\left[\frac{8(f_0 t)^2}{2N+1}\right]\exp\left[\frac{8(f_0 t)^2}{2N+1}\right]$

$\boxed{(N+1)/(2N+1) \cong \tfrac{1}{2}} \to = \exp\left[\frac{(N+1)(4f_0 t)^2}{(2N+1)^2}\right] \cong \exp[(2f_0 t)^2/(N+\tfrac{1}{2})]. \qquad (A3)$

We conclude that, for sufficiently large $N$, the truncated Euler expansion of Eq.(A1) provides a good approximation to $\cos(2\pi f_0 t)$ wherever the coefficient $\exp[(2f_0 t)^2/(N+\tfrac{1}{2})]$ comes reasonably close to 1. This occurs over the time interval $|t| \lesssim \sqrt{N}/(4f_0)$, where the value of the coefficient falls between 1 and $e^{1/4} \cong 1.28$.

## Appendix B

When convolving the functions $f(x')$ and $g(x')$, we hold one of the functions, say, $f(x')$, fixed, and flip the other one around the vertical axis to get $g(-x')$. We then shift the flipped function by a distance $x$ along the horizontal axis to get $g(x-x')$. At this point, we multiply the two functions together and integrate the product over the entire $x'$-axis to obtain $f(x) * g(x) = \int_{-\infty}^{\infty} f(x')g(x-x')dx'$, which, by definition, is the convolution of the two functions evaluated at $x$. By repeating the process for all values of $x$, we build up the desired convolution as a function of $x$.

Applying the above procedure to a pair of unit-width rectangular functions, $f(x) = g(x) = \text{rect}(x)$, we find that the processes of shifting, multiplying, and integrating eliminate the discontinuities of the original functions. The end result of the convolution operation is thus the continuous triangular function $\mathrm{tri}(x) = (1 - |x|)\mathrm{rect}(x/2)$. This function, which is twice as wide as our original rectangular functions, is continuous, albeit with a discontinuous first derivative at $x = 0$ and $x = \pm 1$. If we now convolve $\mathrm{tri}(x)$ with $\mathrm{rect}(x)$, we will get a function that is not only continuous and differentiable, but also has a continuous first derivative; the discontinuities now show up in the second derivative. The process can be repeated any number of times, and each time one more derivative becomes continuous.

If we represent each discontinuity with a step-function, then, in a convolution integral, the step-function appears under the integral sign as $\mathrm{step}(x - x')$. The first derivative of the convolution then resembles the same integral, albeit with the step-function replaced by the Dirac delta-function $\delta(x - x')$. The sifting property of the $\delta$-function ensures that, if the other function under the integral sign is continuous, then the derivative of the convolution will be continuous as well. Proof by induction now confirms that, if the other function under the integral sign is $n$ times differentiable, the convolution will be $n + 1$ times differentiable. Here, we need to invoke the sifting property of the $n^{\text{th}}$ derivative of the $\delta$-function, which sifts out the $n^{\text{th}}$ local derivative of the accompanying function under the integral sign.

A function such as $\mathrm{sinc}(t)$ is readily seen to be square-integrable because its Fourier transform $\mathrm{rect}(f)$ is square-integrable. However, the function $t\,\mathrm{sinc}(t)$ is *not* square-integrable because its Fourier transform, which is proportional to the derivative of $\mathrm{rect}(f)$, consists of a pair of $\delta$-functions at $f = \pm\frac{1}{2}$, and these $\delta$-functions are *not* square-integrable. In a similar vein, the function $\mathrm{sinc}^2(t)$, whose Fourier transform is $\mathrm{tri}(f)$, is square-integrable, as is the function $t\,\mathrm{sinc}^2(t)$, whose Fourier transform is proportional to the first derivative of $\mathrm{tri}(f)$. However, the function $t^2\mathrm{sinc}^2(t)$ is *not* square-integrable, since its Fourier transform, being proportional to the second derivative of $\mathrm{tri}(f)$, contains $\delta$-functions, which are *not* square-integrable.

It is not difficult now to extend the above argument to $\mathrm{sinc}^\nu(t)$, where $\nu$ is an arbitrary positive integer, and argue that the Fourier transform of $\mathrm{sinc}^\nu(t)$ can be differentiated $\nu$ times until $\delta$-functions appear in the frequency domain. Since the Fourier transform of the product of the polynomial function $\sum_{n=0}^{N} a_n t^n$ and $\mathrm{sinc}^\nu(t)$ is a linear superposition of the Fourier transform of $\mathrm{sinc}^\nu(t)$ and its various derivatives up to the order $N$, we conclude that the product function $(\sum_{n=0}^{N} a_n t^n)\mathrm{sinc}^\nu(t)$ will be square-integrable provide that $\nu > N$.

In general, one can state that the function $g(t)$, whose Fourier transform $G(f)$ is bandlimited, square-integrable, continuous, and $N$ times differentiable before the appearance of $\delta$-functions in its derivatives, will remain bandlimited and square-integrable when multiplied by the $N^{\text{th}}$ order polynomial function $\sum_{n=0}^{N} a_n t^n$.

## Appendix C

This appendix aims to confirm that the functions $g(t)$, listed in paragraphs (i), (ii), (iii) of Sec. 3, are indeed derived from their corresponding $G(f)$ via the inverse Fourier integral $g(t) = \int_{-\infty}^{\infty} G(f)\exp(\mathrm{i}2\pi ft)\,\mathrm{d}f$. In the case of the function $G(f)$ of paragraph (i) we have

$$\int_{-1}^{1} \tfrac{3}{4}(1 - f^2)\exp(\mathrm{i}2\pi ft)\,\mathrm{d}f = \tfrac{3}{2}\mathrm{sinc}(2t) + \frac{3\,\mathrm{sinc}''(2t)}{8\pi^2} = \frac{3[\mathrm{sinc}(2t) - \cos(2\pi t)]}{(2\pi t)^2} = g(t). \quad (\text{C1})$$

The peak of the above function at $g(0) = 1$ is consistent with the fact that $\int_{-1}^{1} G(f)\mathrm{d}f = 1$. In the case of the function $G(f)$ of paragraph (ii), Sec. 3, we resort to the table of integrals (G&R **3.387**-2)[36] to write $\int_{-1}^{1}(1-f^2)^\kappa \exp(i2\pi ft)\,df = \sqrt{\pi}\,\Gamma(\kappa+1)J_{\kappa+\frac{1}{2}}(2\pi t)/(\pi t)^{\kappa+\frac{1}{2}} = g(t); \quad [\text{Re}(\kappa)>-1].$ (C2)

Here, $J_{\kappa+\frac{1}{2}}(\cdot)$ is the Bessel function of first kind, order $\kappa+\frac{1}{2}$. The peak value of $g(t)$ of Eq.(C2) occurs at $t=0$, where $g(0)=\sqrt{\pi}\,\Gamma(\kappa+1)/\Gamma(\kappa+\tfrac{3}{2})$; see G&R **8.440**.[36] In the case of $G(f)$ of paragraph (iii), Sec.3, we find (G&R **3.892**-2)[36]

$\int_{-\pi/2}^{\pi/2}\cos^\kappa(f)\exp(i2\pi ft)\,df = \dfrac{\pi}{2^\kappa(\kappa+1)B[1+\frac{1}{2}\kappa+\pi t,\ 1+\frac{1}{2}\kappa-\pi t]} = g(t); \quad [\text{Re}(\kappa)>-2].$ (C3)

The Euler beta function $B(\cdot,\cdot)$ appearing in the preceding equation is related to the gamma function via $B(x,y)=\Gamma(x)\Gamma(y)/\Gamma(x+y)$; see G&R **8.384**-1.[36] Thus, the function $g(t)$ of Eq.(C3) can be further streamlined to yield

$$g(t) = \dfrac{\pi\Gamma(\kappa+1)}{2^\kappa\Gamma(1+\frac{1}{2}\kappa+\pi t)\,\Gamma(1+\frac{1}{2}\kappa-\pi t)}.$$ (C4)

The peak value of the above $g(t)$ occurs at $t=0$, where $g(0)=\sqrt{\pi}\,\Gamma(\frac{1}{2}+\frac{1}{2}\kappa)/\Gamma(1+\frac{1}{2}\kappa)$; see G&R **8.331**-1 and **8.335**-1.[36]

## Appendix D

Let $g(t)$ be a square-integrable and bandlimited function whose Fourier transform $G(f) = \int_{-\infty}^{\infty} g(t)\exp(-i2\pi ft)\,dt$ equals zero when $|f|>f_0$. We prove that any zero of $g(t)$, say, the one at $t=t_1$, can be removed upon dividing $g(t)$ by $t-t_1$ without increasing the original bandwidth. The Fourier transform of $1/(t-t_1)$ is readily evaluated as follows:

$\mathcal{F}\{1/(t-t_1)\} = \lim_{\varepsilon\to 0}\left\{\int_{-\infty}^{t_1-\varepsilon}\dfrac{\exp(-i2\pi ft)}{t-t_1}\,dt + \int_{t_1+\varepsilon}^{\infty}\dfrac{\exp(-i2\pi ft)}{t-t_1}\,dt\right\}$

$= -i\pi\,\text{sign}(f)\exp(-i2\pi ft_1).$ (D1)

Alternatively, one may evaluate the Fourier integral $\int_{-\infty}^{\infty}[\exp(-i2\pi ft)/(t-t_1)]\,dt$ in the complex $z$-plane, where $z=t_1$ is the sole singularity of the integrand. The semi-circular contour of integration for $f<0$ must be in the upper-half, and for $f>0$ in the lower-half, of the $z$-plane. In both cases, the residue is $\exp(-i2\pi ft_1)$, which must then be multiplied by $i\pi$ when $f<0$, and by $-i\pi$ when $f>0$. The Fourier transform is seen to be that given in Eq.(D1).

Next, consider the function $\tilde{g}(t) = g(t)/(t-t_1)$, whose Fourier transform is obtained by a convolution between $G(f)$ and the Fourier transform of $1/(t-t_1)$ given by Eq.(D1), namely,

$\tilde{G}(f) = G(f) * [-i\pi\,\text{sign}(f)\exp(-i2\pi ft_1)]$

$= -i\pi \int_{-\infty}^{\infty} G(f')\text{sign}(f-f')\exp[-i2\pi(f-f')t_1]\,df'$

$= -i\pi\exp(-i2\pi ft_1)\int_{-f_0}^{f_0} G(f')\text{sign}(f-f')\exp(i2\pi t_1 f')\,df'.$ (D2)

For $|f|>f_0$, the function $\text{sign}(f-f')$ appearing in Eq.(D2) is constant (either $+1$ or $-1$) within the integration domain $[-f_0,f_0]$. Consequently,

$\tilde{G}(f) = -i\pi\,\text{sign}(f)\exp(-i2\pi ft_1)\int_{-f_0}^{f_0} G(f')\exp(i2\pi t_1 f')\,df'$

$= -i\pi\,\text{sign}(f)\exp(-i2\pi ft_1)\,g(t_1) = 0;\qquad |f|>f_0.$ (D3)

It is thus seen that, when $g(t_1) = 0$, the bandwidth of $\tilde{G}(f)$, the spectrum of $\tilde{g}(t) = g(t)/(t - t_1)$, must be less than or equal to that of $G(f)$.

Higher-order zeros of $g(t)$ can be similarly removed by repeated application of the above procedure, i.e., by first forming $\tilde{g}(t) = g(t)/(t - t_1)$, then applying the procedure to $\tilde{g}(t)$ to form $\tilde{\tilde{g}}(t) = \tilde{g}(t)/(t - t_1)$, and so on. In other words, if $t_1$ happens to be an $n^{\text{th}}$ order zero of the square-integrable, bandlimited function $g(t)$, then the bandwidth of $g(t)/(t - t_1)^n$ cannot exceed that of the original spectrum $G(f)$.